\documentclass[aps,prl,twocolumn,letterpaper,superscriptaddress,showpacs]{revtex4}
\usepackage{graphicx}
\usepackage{CJK}

\begin{document}
\begin{CJK*}{UTF8}{bsmi}
\title{
Effective doping and suppression of Fermi surface reconstruction via Fe vacancy disorder in K$_x$Fe$_{2-y}$Se$_2$
}
\author{Tom Berlijn}
\affiliation{Condensed Matter Physics and Materials Science Department,
Brookhaven National Laboratory, Upton, New York 11973, USA}
\author{P. J. Hirschfeld}
\affiliation{Department of Physics, University of Florida}
\author{Wei Ku(%
顧威)
}
\affiliation{Condensed Matter Physics and Materials Science Department,
Brookhaven National Laboratory, Upton, New York 11973, USA}
\affiliation{Physics Department, State University of New York, Stony Brook,
New York 11790, USA}

\date{\today}

\begin{abstract}
We investigate the effect of disordered vacancies on the normal-state electronic structure of the newly discovered alkali-intercalated iron selenide superconductors.
To this end we use a recently developed Wannier function based method to calculate from first principles the configuration-averaged spectral function $\langle A(k,\omega)\rangle$ of K$_{0.8}$Fe$_{1.6}$Se$_2$ with disordered Fe and K vacancies.
We find that the disorder can suppress the expected Fermi surface reconstruction without completely destroying the Fermi surface.
More interestingly, the disorder effect raises the chemical potential significantly, giving enlarged electron pockets almost identical to highly doped KFe$_2$Se$_2$, without adding carriers to the system.
\end{abstract}


\maketitle
\end{CJK*}

The newly discovered ~\cite{guo} alkali-intercalated iron selenide selenide superconductors A$_x$Fe$_{2-y}$Se$_2$, with A=K, Rb, Cs, share a set of properties that distinguish them  from any of the other iron based superconductor(FeSC) families.
Their transition temperature is found to be as high as $T_c=31$K at ambient pressure ~\cite{guo}, high in comparison with the other 11-type chalgogenides.
They are the first FeSC on the verge of becoming an antiferromagnetic (AFM) insulator~\cite{mhfang}.
The (AFM) order has a novel block type structure with an unprecedented high transition temperature of $T_N=559$K and magnetic moment of 3.31$\mu_B$/Fe ~\cite{bao}.
From ARPES experiments ~\cite{xpwang,zhang,zhao,mou,qian,chen,borisenko}, only electron pockets are found and no hole pockets, seemingly ruling out the popular $s^{\pm}$ pairing scenario \cite{mazin}.  Last, but from an electronic structure point of view certainly not least, they contain significant numbers ($\sim 20\%$) of Fe and K vacancies (V$_{\rm Fe}$ and V$_{\rm K}$).

While the presence of these vacancies has been well accepted, their distribution is currently under debate.
From the superlattice peaks in neutron diffraction studies ~\cite{bao} it is concluded that the Fe vacancies form a $\sqrt{5}\times\sqrt{5}$ superlattice with strong long-range order.
On the other hand, a different scenario has been suggested for those samples which display simultaneously long range magnetic order and superconductivity,
namely that of a phase separation, whereby the ordered Fe vacancies only exist in one of the phases.
Some scanning tunneling microscopy \cite{li}, nuclear magnetic resonance \cite{texier}, and transmission electron microscopy (TEM) measurements~\cite{yan} conclude that the second phase contains no Fe vacancies.
Other TEM ~\cite{zwang,chen} and x-ray absorption near edge structure measurements ~\cite{iadecola} conclude that the second phase contains disordered Fe vacancies.

An obvious puzzle emerges immediately upon examining angle-resolved photoemission spectroscopy (ARPES) measurements~\cite{xpwang,zhang,zhao,mou,qian,chen, borisenko}.
As was demonstrated theoretically~\cite{lin}, the $\sqrt{5}\times\sqrt{5}$ superlattice of Fe vacancies must induce a strong reconstruction of the Fermi surface and the band structure, due to the huge scattering associated with the vacancy.
However, current ARPES results all show a simple electronic structure without any indication of reconstruction.
A possible solution~\cite{lin} to this puzzle is that the regions to which ARPES measurements are sensitive must contain poorly ordered Fe vacancies.
This explanation, however, is vulnerable to the potential problem that while vacancy disorder might suppress the band structure reconstruction, it may also destroy completely the Fermi surface.  
It is thus timely to examine the physical effects of Fe vacancy disorder in this class of materials.

The ARPES experiments also reveal a peculiar behavior of the A$_x$Fe$_{2-y}$Se$_2$ family.
Regardless of the nominal value of $x$ and $y$, the observed Fermi surfaces always contains large electron pockets in the zone corner, as if the system is strongly electron doped compared to the ``245'' parent compound A$_{0.8}$Fe$_{1.6}$Se$_2$.
In fact, the chemical potential is shifted so much that not only do the hole pockets at the zone center submerge entirely below the chemical potential, but also a small electron pocket appears instead.
Consequently, these systems are commonly regarded as strongly electron doped, closer to A$_1$Fe$_2$Se$_2$ (122) than 245, without a reasonable account of the origin of the large number of extra electrons.
Considering the essential role of doping in the magnetism and superconductivity of these materials, a physical explanation of this peculiar behavior is clearly of highest importance.

In this Letter, we address these two important issues by studying the influence of vacancy disorder on the alkali-intercalated Fe selenides, using a recently developed Wannier function based method to calculate from first principles the configuration-averaged spectral function $\langle A(k,\omega)\rangle$ of K$_{0.8}$Fe$_{1.6}$Se$_2$ with disordered Fe and K vacancies.
We find that the disorder does indeed suppress the band structure reconstruction without completely destroying the Fermi surface.
Furthermore, we find that the disorder  raises  the chemical potential significantly, eliminating the hole pockets and enlarging the electron pockets, in excellent agreement with the ARPES experiments, \textit{without} adding carriers.
We provide a microscopic explanation for the origin of the effective doping in terms of the disorder induced frequency broadenings of the bands.
In addition, we find the  emergence of strongly incoherent carriers.
Our result demonstrates clearly the general inapplicability of Luttinger theorem in disordered systems, and suggests an alternative interpretation of existing ARPES measurements. 

\begin{figure*}[t]
\centering
\includegraphics[width=2\columnwidth,clip=true]{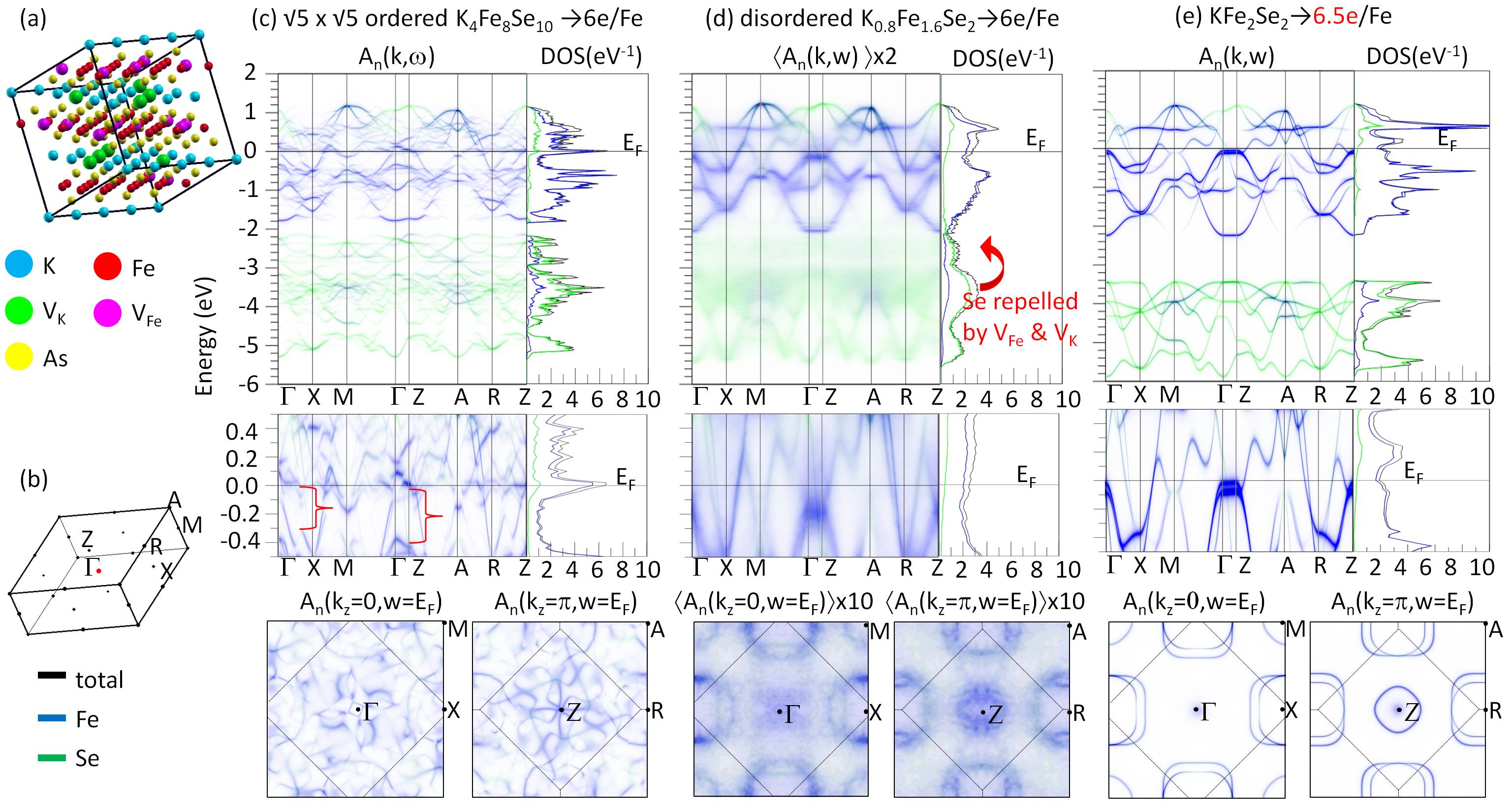}
\caption{
(color online)
(a) An example of a large sized supercell used for the configurational average and (b) 1-Fe Brillouin zone.
Band structure, density of states and Fermi surface of $\sqrt{5}\times\sqrt{5}$ ordered Fe and K vacancies(c), disordered K$_{0.8}$Fe$_{1.6}$Se$_2$(d) and KFe$_2$Se$_2$(d).
The dashed lines in the Fermi surface plots mark the 2-Fe Brillouin zone boundaries.
The band structure and Fermi surface of the disordered K$_{0.8}$Fe$_{1.6}$Se$_2$  are enhanced by a factor of 2 and 10 respectively for better visualization.
}
\label{fig:fig1}
\end{figure*}

The band structure of a disordered system is given by the configuration-averaged spectral function:
$\langle A_n(k,\omega)\rangle$=$\sum_{c}P^{c}A^{c}_n(k,\omega)$ of Wannier orbital~\cite{wannier} $n$, crystal momentum $k$ and frequency $\omega$,
in which configuration $c$ is weighted by its probability $P^c$.
By treating the disordered configurations within the supercell approximation, their spectral functions $A^{c}_n(k,\omega)$
can be obtained directly from the supercell eigenvectors and eigenvalues, using the unfolding method~\cite{unfolding}.
To handle the computational cost related to the large sizes of the supercells, essential for a proper treatment of the disorder, we employ the recently developed Wannier function based effective Hamiltonian method~\cite{naxco2}.
The low energy Hilbert space is taken in [-6,2]eV consisting of the Wannier orbitals of Fe-$d$ and Se-$p$ characters.
The influence of the Fe and K vacancies is extracted from three DFT~\cite{sup} calculations: the clean KFe$_2$Se$_2$ and the single vacancy supercells
K$_7$Fe$_{16}$Se$_{16}$ and K$_8$Fe$_{15}$Se$_{16}$.
For the configurational average, we use 10 large supercells (e.g. Fig.\ref{fig:fig1}(b)) of random size, shape and orientation containing 176 atoms on average, and assume an equal probability among the configurations.
 Benchmarks demonstrating the high accuracy of the effective Hamiltonian method and the convergence of the configuration average against the size and the number of configurations are given in Ref.~\cite{sup}.
 
As a reference, let's first re-examine the fully $\sqrt{5}\times\sqrt{5}$ ordered 245 system that shows strong band structure reconstruction~\cite{lin}.
To facilitate the comparison with ARPES, Figure~\ref{fig:fig1}(c) shows the band structures and Fermi surfaces in the 1-Fe Brillouin zone (c.f. Fig.~\ref{fig:fig1}(a)).
Indeed, the lower panel shows a very strong reconstruction of the Fermi surface that bears no resemblance to the observed ARPES measurements ~\cite{xpwang,zhang,zhao,mou,qian,chen, borisenko}.
This strong Fermi surface reconstruction is not a special property of the $\sqrt{5}\times\sqrt{5}$ ordering, but expected to happen for other vacancy orderings~\cite{yan,zwang} in general, due to the very strong scattering against the Fe vacancies.
The top panel also shows that the whole band structure, not just the Fermi surface, is reconstructed, with gap openings and complex folded features occurring everywhere in both Fe and Se bands.
For example, the red bars in the middle panel indicate two large gap opening on the order of 300-400 meV.
These kinds of strong reconstructions should in principle allow ARPES to quantify the strength of the vacancy order, including the semiconducting samples that exhibit no Fermi surface.
The fact that no such reconstruction has been reported is at odds with the strong vacancy order observed by neutron scattering~\cite{bao}.

The configuration-averaged spectral function in Fig.~\ref{fig:fig1}(d) confirms that  strong disorder in the vacancies introduces very large incoherent scatterings.
The Fe bands now develop large linewidths in both momentum and frequency, reflecting their short mean free path and lifetime.
More dramatically, the top of the Se bands within [-4,-2]eV even becomes unrecognizable, due to the large repulsive potential of $\sim$0.6eV/0.2eV from Fe/K vacancies on their 4/8 neighboring Se orbitals.
Nevertheless, even for such a completely disordered case, Fig.~\ref{fig:fig1}(d) shows that the Fe band structure and the Fermi surface survive the strong vacancy disorder effects, displaying the characteristics of a dispersive band structure.
On the other hand, Fig.~\ref{fig:fig1}(d) shows the clear suppression of the reconstruction in the band structure and the Fermi surface.
All the gap openings and folded features are washed out.
In fact, other than the strong broadening effects and a clear reduction of bandwidth, the resulting band structure and Fermi surfaces agree very well with those of the clean 122 system in Fig.~\ref{fig:fig1}(e).
We have thus provided quantitative support to the notion that the regions to which ARPES measurements are sensitive must contain poorly ordered Fe vacancies.

Figure~\ref{fig:fig1}(d) also shows an unexpected effect of the vacancy disorder, namely a large shift of the chemical potential.
Indeed, the chemical potential is now $\sim$200meV above the tip of the hole pockets, and allows a small electron pocket to appear around the Z point.
Correspondingly, the electron pockets at the X/R point are much larger than the parent compounds of all families of Fe-based superconductors.
In fact, the resulting Fermi surface looks exactly like that of the heavily doped 122 system in Fig.~\ref{fig:fig1}(e), in excellent agreement with the ARPES measurements, even though the system actually contains \textit{no} additional electrons.
(A Luttinger count of the carrier density from the volume of the Fermi surface would give the illusion of +0.5/Fe doping for this undoped system.)
This is a clear demonstration that Luttinger theorem should never be applied to systems with strong disorder~\cite{haverkort, tm122}.
Specifically in this case, our result suggests that the experimental samples showing 122 like Fermi surface might not even be doped at all.

\begin{figure}
\includegraphics[width=0.9\columnwidth,clip=true]{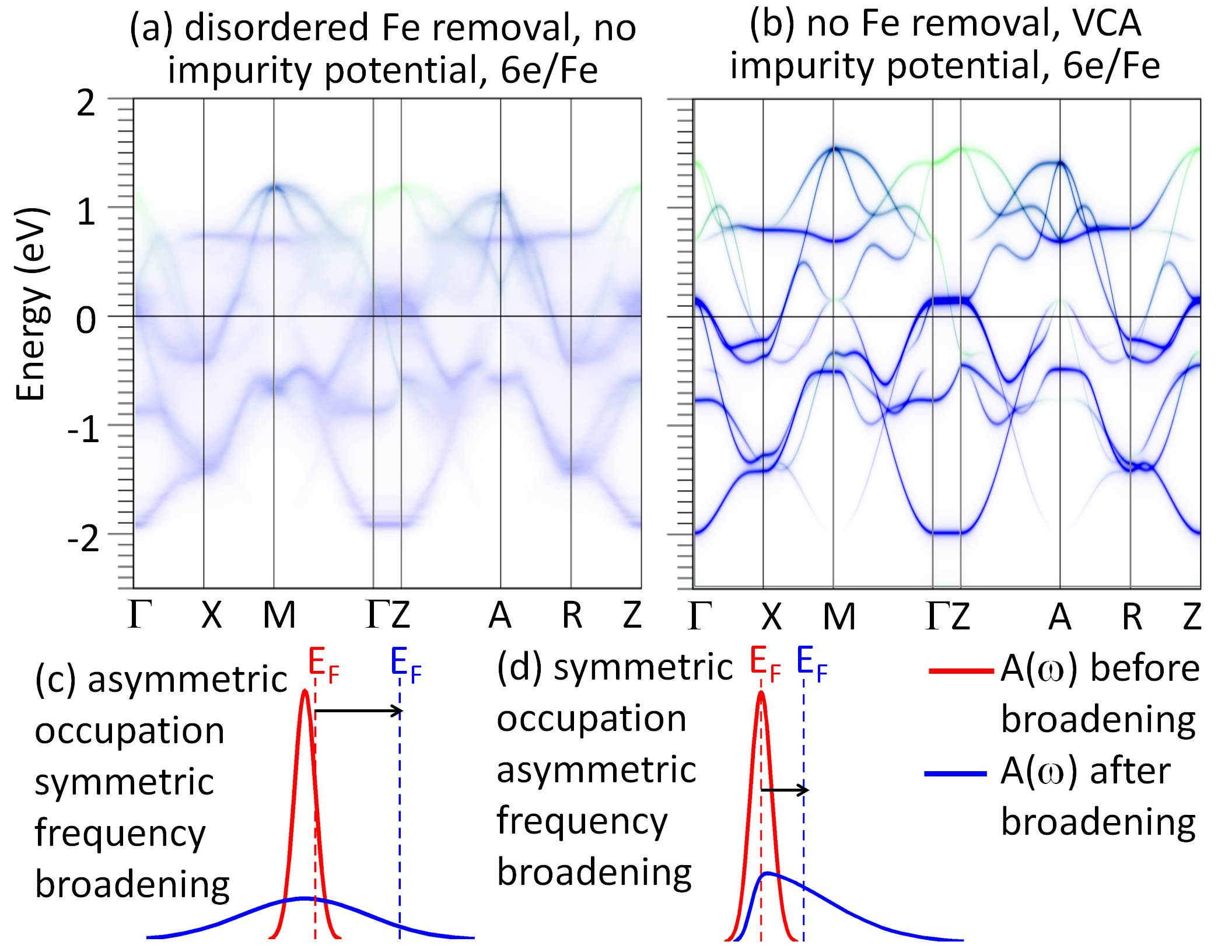}
\caption{\label{fig:fig2}
(color online)
Spectral function obtained from disordered removal of Fe orbitals alone (without the impurity potentials) (a), and from the virtual crystal of the impurity potentials (without the removal of Fe orbitals).
Both show negligible effective doping in comparison with the full disordered calculation.
(c)(d) Illustration of microscopic mechanisms contributing to effective doping through broadening in frequency.
}\end{figure}

Microscopically, this strong shift of chemical potential (or an \textit{effective} ``doping'') is to be distinguished from real doping, for example in the case of Co substituted BaFe$_2$As$_2$, in which the chemical potential shift is a direct consequence of the additional valence electron in the system~\cite{tm122}. (This conclusion differs from the previous studies~\cite{nakamura, konbu}).
Figure~\ref{fig:fig2}(a) reveals more insight by ignoring the impurity potential in the calculation (but still removing Fe orbitals at the disordered vacancy sites).
In this case, the effective doping becomes negligible, as an ``expected'' chemical potential of an undoped system is found, giving both hole and electron pockets of regular sizes.
On the other hand, an average account of the impurity potential via the virtual crystal approximation (c.f. Fig.~\ref{fig:fig2}(b)) misses entirely the effective doping as well.
Therefore, the strong chemical potential shift is only realized from the disordered nature of the impurity potential.

Figure~\ref{fig:fig2} illustrates two microscopic mechanisms that contribute to the effective doping through disorder induced broadenings in frequency.
For simplicity, let's consider the broadening of the momentum independent spectral function $A(\omega)$.
First, in Fig.~\ref{fig:fig2}(c), when the spectral weight near the chemical potential (within the frequency range of the broadening) is larger below, a symmetric broadening in frequency pushes the chemical potential up.
Second, in Fig.~\ref{fig:fig2}(d), an asymmetric broadening skewed toward higher frequency alone does the same\cite{mechanisms}.
The above test cases in Fig.~\ref{fig:fig2}(a) and (b) suggest that the essential mechanism in K$_x$Fe$_{2-y}$Se$_2$ is the aymmetric broadening resulting from the repulsive impurity potential.

Without increasing the total electronic density of the system, the significant shift of chemical potential leads to many important physical consequences through, for example, the change of Fermi velocity and the Fermi wave vector, in a manner similar to real doping in regular systems without disorder.
In particular, the significant suppression of phase space for long-range magnetic order around ($\pi$,0) resembles very much the effect of real doping.
This offers a natural explanation of the occurrence of a superconducting phase in this undoped system, and suggests a novel mechanism of promoting superconductivity via strongly disordered impurity potentials.

Interestingly, there have indeed been experimental observations suggesting enhancement of superconductivity via disorder in these systems.
Reference~\cite{han} reported that the same sample can be made superconducting/magnetically ordered through quenching/annealing.
High pressure experiments~\cite{lsun} reported a seemingly second superconducting transition at considerably higher temperature $T_c=48K$ at pressure above 11.5GPa, consistent with promotion of superconductivity via pressure-induced suppression of vacancy order.
A second superconducting phase has also been recently reported at ambient pressure ~\cite{amzhang2}.

Obviously, for disordered impurities to promote superconductivity, their positive influence (from the above effective doping or other disorder effects discussed below) must overcome the negative influence of disorder (for example pair-breaking or phase fluctuation scattering).
This issue is intimately tied to the nature of the scatterer (for example, magnetic or nonmagnetic, weak scattering or strong scattering regime, and so on) and the superconductivity itself (for example, sign changing order parameter or not, dominant amplitude fluctuation or phase fluctuation, and so on).
We anticipate rich physics to be revealed from future quantitative investigations along this direction.

\begin{figure}
\includegraphics[width=1\columnwidth,clip=true]{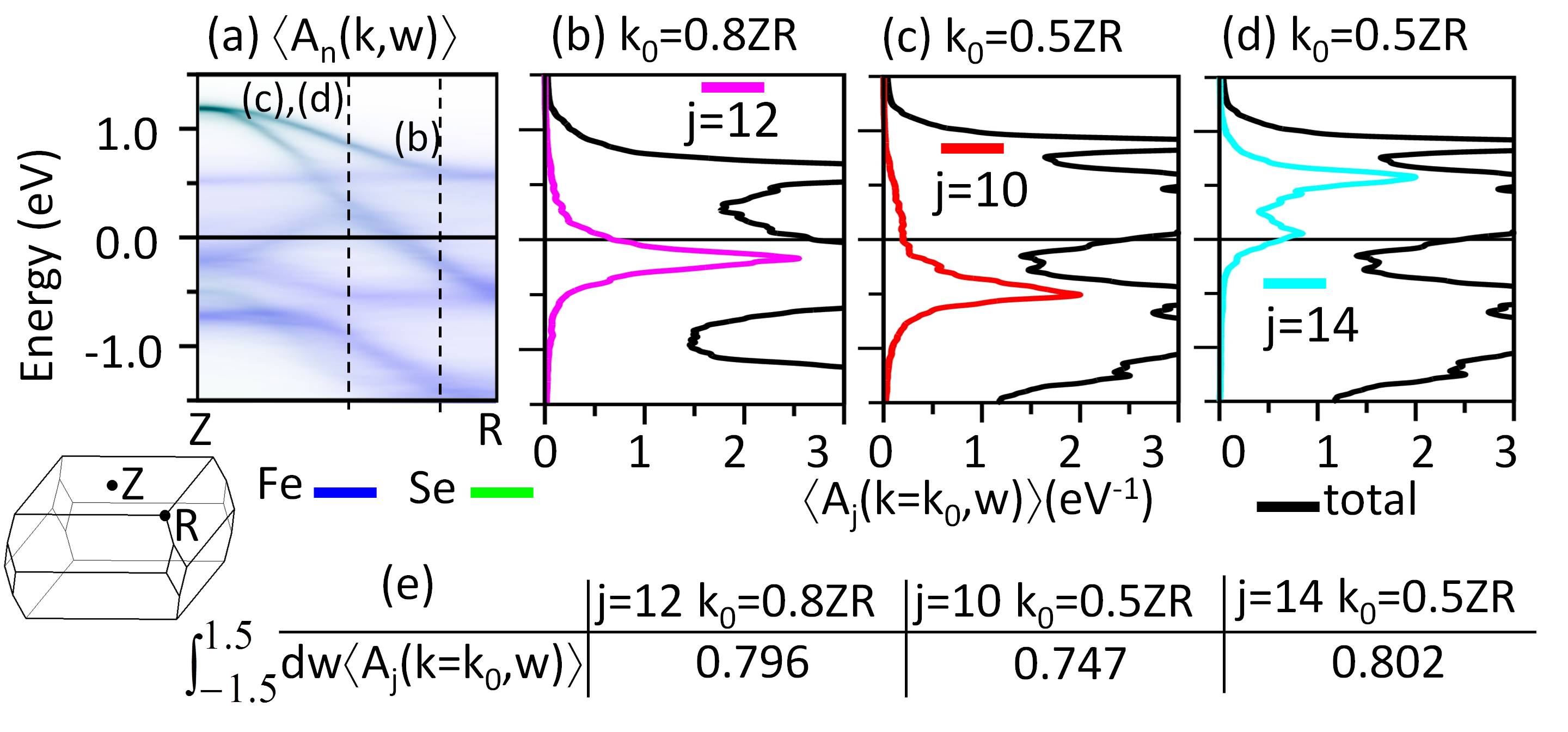}
\caption{\label{fig:fig3}
(color online)
Disorder induced spectral weight distributions, represented in the 2-Fe Brillouin zone,
(a) as a function of crystal momentum $k$ and orbital indices $n$, 
and at fixed crystal momenta $k_0$ and band index $j$ with
(b) $k_0=0.8$ZR and $j=12$ belonging to the large electron pocket in the zone corner,
(c) $k_0=0.5$ZR and $j=10$ belonging to the hole pocket leaking into the Fermi surface,
(d) $k_0=0.5$ZR and $j=14$ displaying an incoherent satellite at the Fermi surface and
(e) their integrations displaying $\sim$20\% loss of spectral weight due to the Fe vacancies.
}\end{figure}

It is also important to account for the removal of Fe orbitals in the vacancy sites.
Indeed, representing the spectral function in the eigenbasis $|kj\rangle$ of crystal momentum $k$ and band index $j$ of the uniform 122 system, $A_{j}(k,\omega)=-\frac{1}{\pi}{\rm Im} \langle kj|G(\omega)|kj\rangle$~\cite{tm122}, Fig.~\ref{fig:fig3} shows that the total spectral weights of all well-defined low-energy quasi-particle drop to $\sim$0.8, in good accordance to the 20\% Fe vacancy.
As advocated previously~\cite{tm122}, such a large reduction in the spectral weight is expected to change strongly (even qualitatively) the standard picture of spin fluctuations~\cite{maier,fwang}, through a significant reduction of the bare susceptibility by roughly a factor of $0.8^2\sim0.6$.
The general argument~\cite{tm122} of suppression/resilience of phase space of antiferromagnetic/superconducting order should remain valid, which leads to a potential promotion of superconductivity.

In addition, our result reveals strong incoherent features in this system.
For example, Fig.~\ref{fig:fig3}(c) illustrates that small amount of diffusive hole carriers in the system are still present in this seemingly highly electron doped system, through the long tails of the hole bands.
This suggest that hole carriers could still contribute to the effective pairing~\cite{fwang}.
Similarly, Fig.~\ref{fig:fig3}(d) shows the presence of highly incoherent carriers right at the chemical potential, originating from the normally unoccupied band 0.5eV above the chemical potential.
All of these strong effects illustrate vividly the rich physics induced by disorder, well beyond a simple modification of the band filling or dispersion.

Finally, it is important to clarify that our case study with fully disordered vacancies cannot correspond exactly to
the real materials, since that would not be consistent with the observed superstructures from diffraction experiments~\cite{bao, yan,zwang,chen}.
It is instead meant to illustrate the general features of poorly ordered regions of the sample, for example, near the surface where diffusion of vacancies is easier, or in the vacancy disordered region of a phase separated sample.
Also, the vacancies are likely to be short range ordered (SRO), given their charged nature.
Furthermore, additional scattering related to short-range magnetic and orbital correlation~\cite{cclee,sliang} should also be present.  
While we do not expect such SROs to  qualitatively change our conclusions, further studies on their effects are desirable.

In conclusion, we studied from first-principles the normal state configuration-averaged spectral function $\langle A(k,\omega)\rangle$ of K$_{0.8}$Fe$_{1.6}$Se$_2$ with disordered vacancies.
We found that strong vacancy disorder does not destroy the band structure and Fermi surface, but is able to suppress the significant band structure reconstruction expected in vacancy ordered systems, thus producing the smooth quasi-particle band structure observed by ARPES.
Furthermore, we found an intriguing disorder-induced ``effective doping'' that enlarges significantly the electron pockets to be comparable to the highly electron doped KFe$_2$Se$_2$, in excellent agreement with ARPES observations.
Our result demonstrates clearly the general inapplicability of Luttinger theorem in disordered systems, and suggests that the region to which ARPES measurements are sensitive might be almost undoped after all, but with strong vacancy disorder.
Finally, the observed effective doping, together with a strong ($\sim$20\%) loss of coherent spectral weight, offers a novel, general mechanism to promote superconductivity in the presence of strong impurity potentials, as indicated by several current experiments.

Work funded by the U S Department of Energy, Office of Basic Energy Sciences DE-AC02-98CH10886 under the CMCSN program.

\begin{widetext}
{\Large\bf Supplementary information}

\section{Details of the density functional theory calculations}\label{sec:secdft}

\begin{figure}[htp]
\includegraphics[width=0.85\columnwidth]{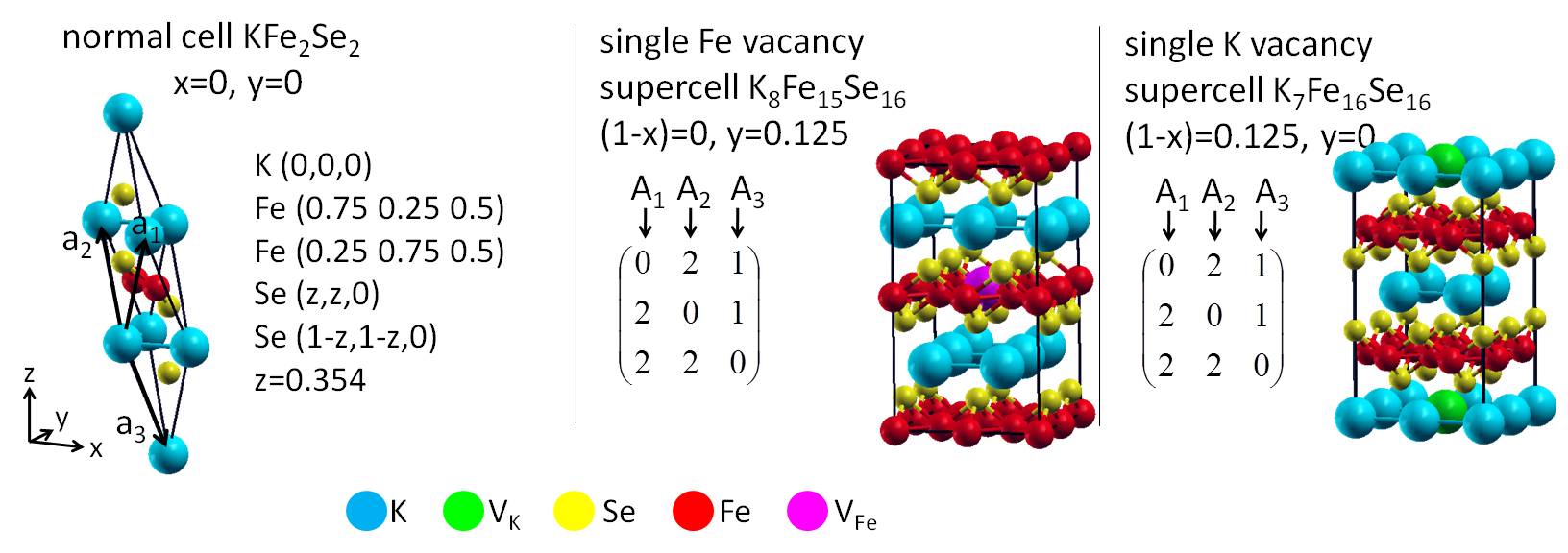}
\caption{\label{fig:figsup1}
(Left) normal cell KFe$_2$Se$_2$, (middle) single Fe vacancy supercell K$_8$Fe$_{15}$Se$_{16}$ and (right) single K vacancy supercell K$_7$Fe$_{15}$Se$_{16}$.}
\end{figure}

The lattice structure of the normal cell, KFe$_2$Se$_2$, is depicted in the left of figure \ref{fig:figsup1}. Its space group I4/mmm and its lattice parameters a=7.46 bohr and c=26.76 bohr and z=0.354, were taken from~\cite{bao1}. The primitive normal cell lattice vectors expressed in Cartesian coordinates are given by:
$ a_1=-\frac{1}{2}a\hat{x}+\frac{1}{2}a\hat{y}+\frac{1}{2}c\hat{z} \; ; \;  a_2=\frac{1}{2}a\hat{x}-\frac{1}{2}a\hat{y}+\frac{1}{2}c\hat{z} \; ; \; a_3=\frac{1}{2}a\hat{x}+\frac{1}{2}a\hat{y}-\frac{1}{2}c\hat{z}$. 
To capture the influence of the Fe and K vacancies an eight times larger supercell is used to properly treat the non-local influence on the nearest As and Fe orbitals. The corresponding K$_8$Fe$_{15}$As$_{16}$/K$_7$Fe$_{16}$Se$_{16}$ supercells are depicted on the right/middle of figure \ref{fig:figsup1} and its primitive lattice vectors expressed in normal cell lattice vectors are given by:
$ A_1=2a_2+2a_3 \; ; \;  A_2=2a_1+2a_3  \; ; \; A_3=a_1+a_2 $. 
We applied the WIEN2K\cite{blaha1} implementation of the full potential linearized augmented plane wave method in the
local density approximation. 
The k-point mesh was taken to be 10$\times$10$\times$10 for the undoped normal cell and 5$\times$5$\times$3 for the supercells respectively.
The basis set sizes were determined by RKmax=7. 

\section{Benchmarks of the effective Hamiltonian against DFT}

To explore the accuracy and efficiency of the effective Hamiltonian method~\cite{naxco21} for the case of K$_x$Fe$_{2-y}$Se$_{2}$, we present comparisons of spectral functions $A_n(k,\omega)$  calculated from the full DFT and the effective Hamiltonian (see figures \ref{fig:figsup2a}-\ref{fig:figsup2c}). The size of the deviations between the full DFT and the effective Hamiltonian should be compared with the size of the impurity induced changes. For this purpose the spectral function of the undoped KFe$_2$Se$_2$ is also plotted as a reference for each benchmark. The test cases in figures \ref{fig:figsup2a} and \ref{fig:figsup2b} are designed as extreme test cases for the ``linearity'' approximation of the impurity influence (see formula (1) of Ref.~\cite{naxco21}). For testing the linearity of the K vacancy influences, the case of Fe$_2$Se$_2$ corresponds to the maximum extrapolation from $(1-x)=0.125$ in the single impurity cell (see right of Fig. \ref{fig:figsup1}) to $(1-x)=1$. For testing the linearity of the Fe vacancy influences, a smaller extrapolation is more meaningful, in order to have remaining Fe bands. The bands of KFeSe$_2$ are very reasonably described, despite the extreme extrapolation in doping from $y=0.125$ in the single impurity cell (see middle of Fig. \ref{fig:figsup1}) to $y=1$. The test case in figure \ref{fig:figsup2c} of a $\sqrt{5}\times\sqrt{5}$ supercell is designed to test the partitioning of the impurity influence from its super images (see section IV of the supplementary information of Ref.~\cite{naxco21}). The basis set of Linear Augmented Plane Waves (LAPW's) used in the full DFT is $\sim30$ times larger then the basis set of Wannier functions used in the effective Hamiltonian method. Since the number of floating point operations of diagonalization depends cubically on the size of the matrix this implies an efficiency increase by a factor of $30^{3}\approx3\cdot10^{4}$. Furthermore the full DFT calculation involves multiple self consistent cycles (typically 15) whereas the effective Hamiltonian method (as currently implemented) requires only a single diagonalization, which increases the efficiency by another order of magnitude to $\sim 6\cdot10^{5}$.

\begin{figure}[htp]
\includegraphics[width=0.8\columnwidth]{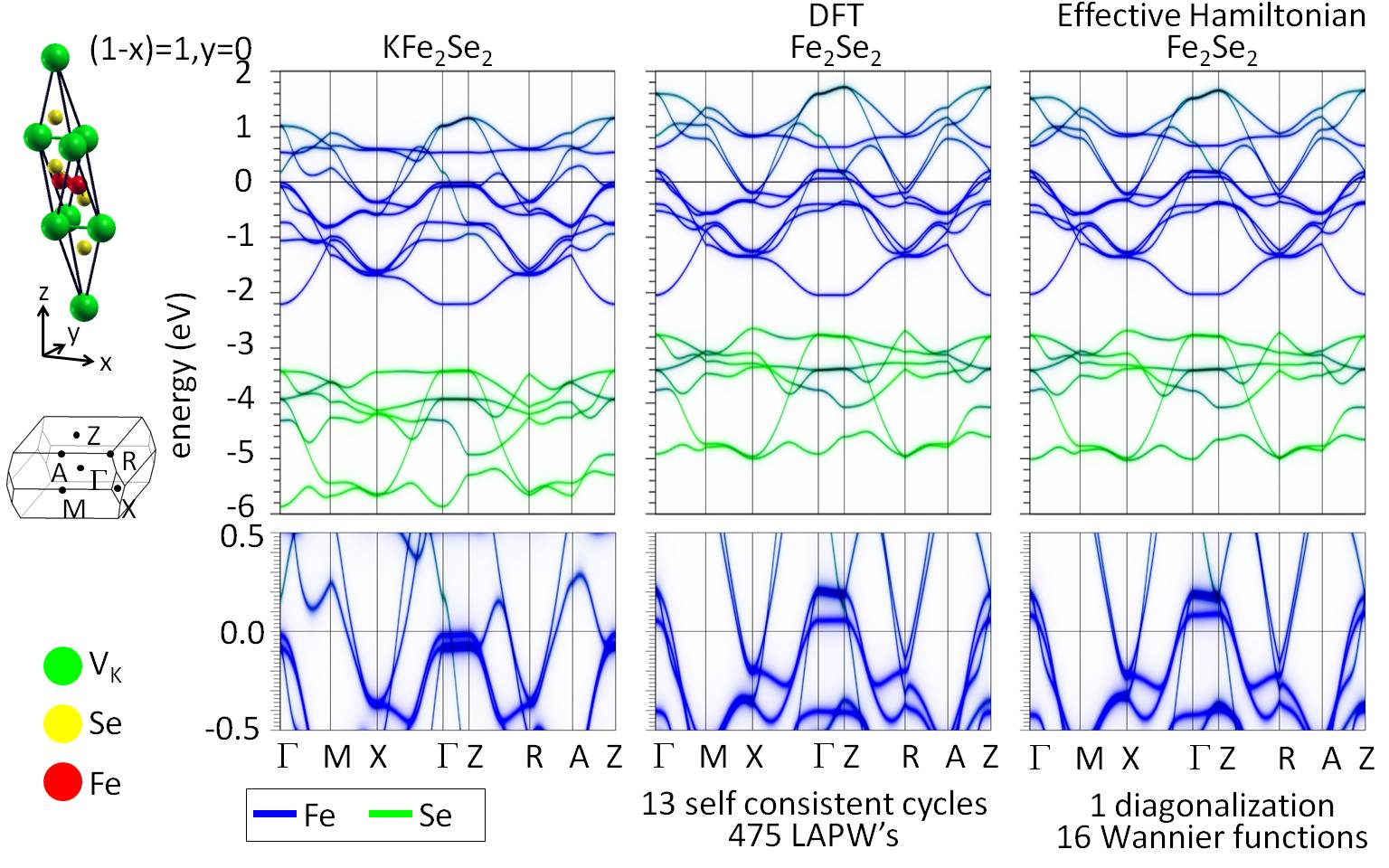}
\caption{\label{fig:figsup2a}
}
\end{figure}

\begin{figure}[htp]
\includegraphics[width=0.8\columnwidth]{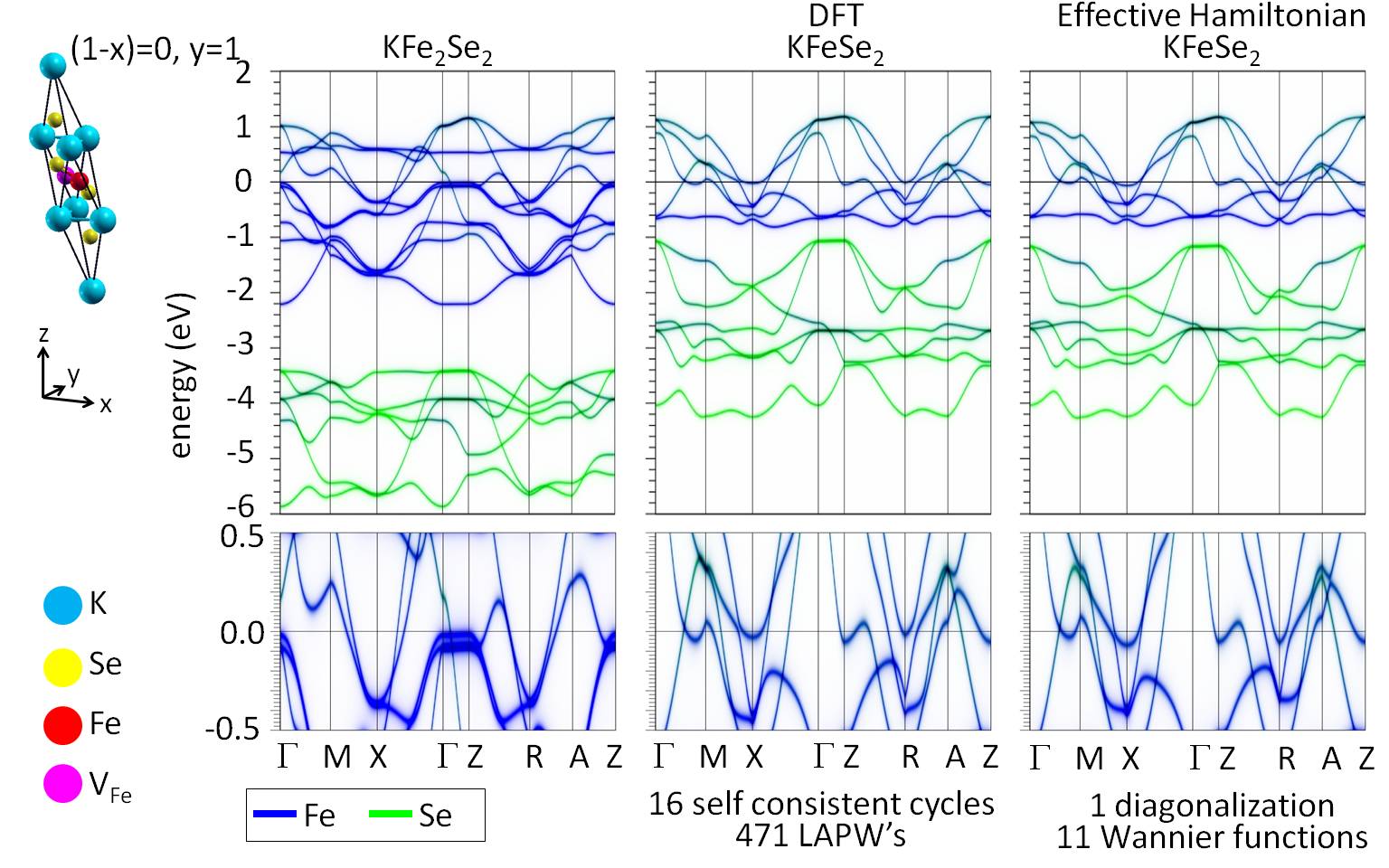}
\caption{\label{fig:figsup2b}
}
\end{figure}

\begin{figure}[htp]
\includegraphics[width=0.8\columnwidth]{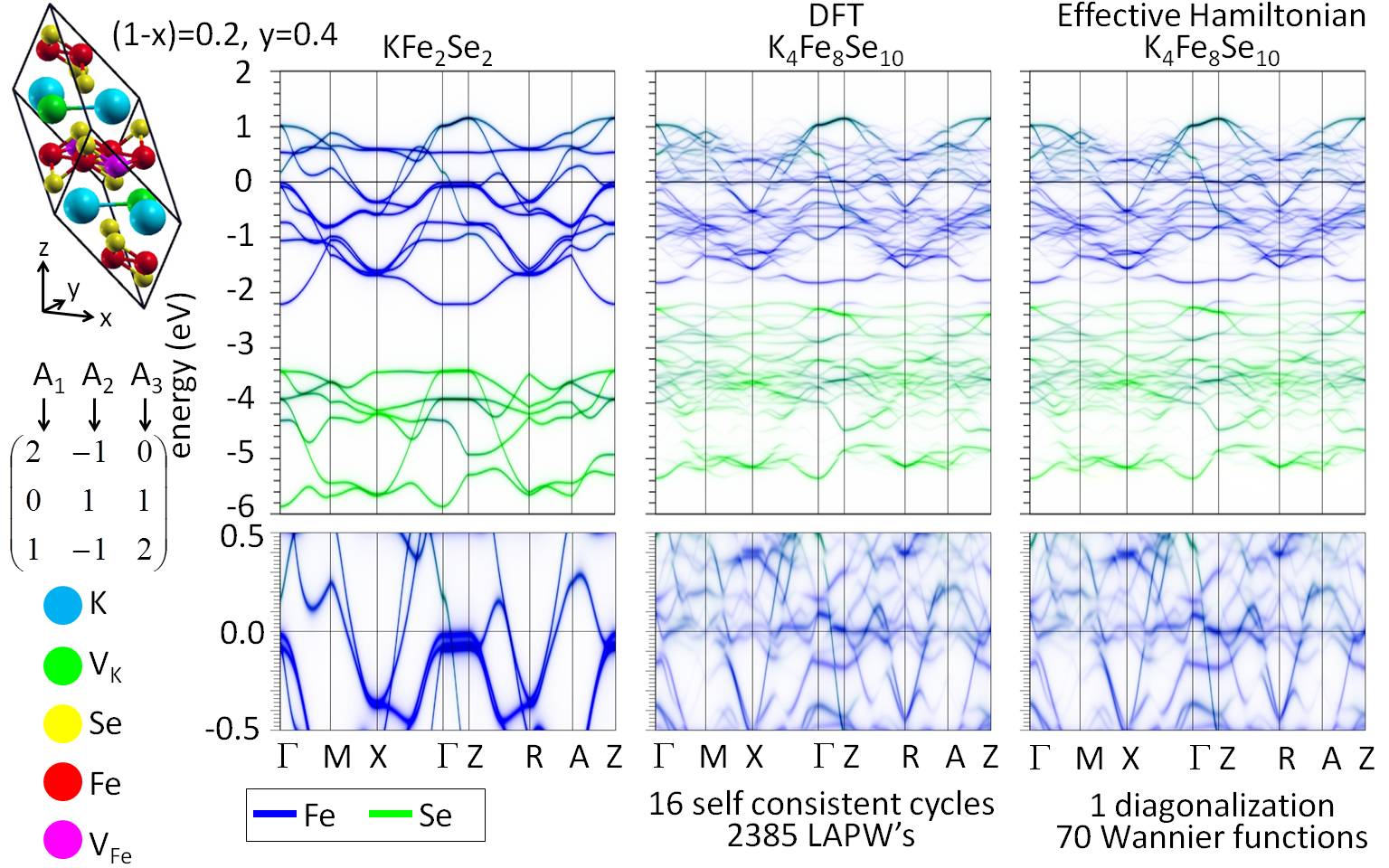}
\caption{\label{fig:figsup2c}
}
\end{figure}

\clearpage
\section{Convergence with respect to size and number of configurations}

\begin{figure}[htp]
\includegraphics[width=0.8\columnwidth]{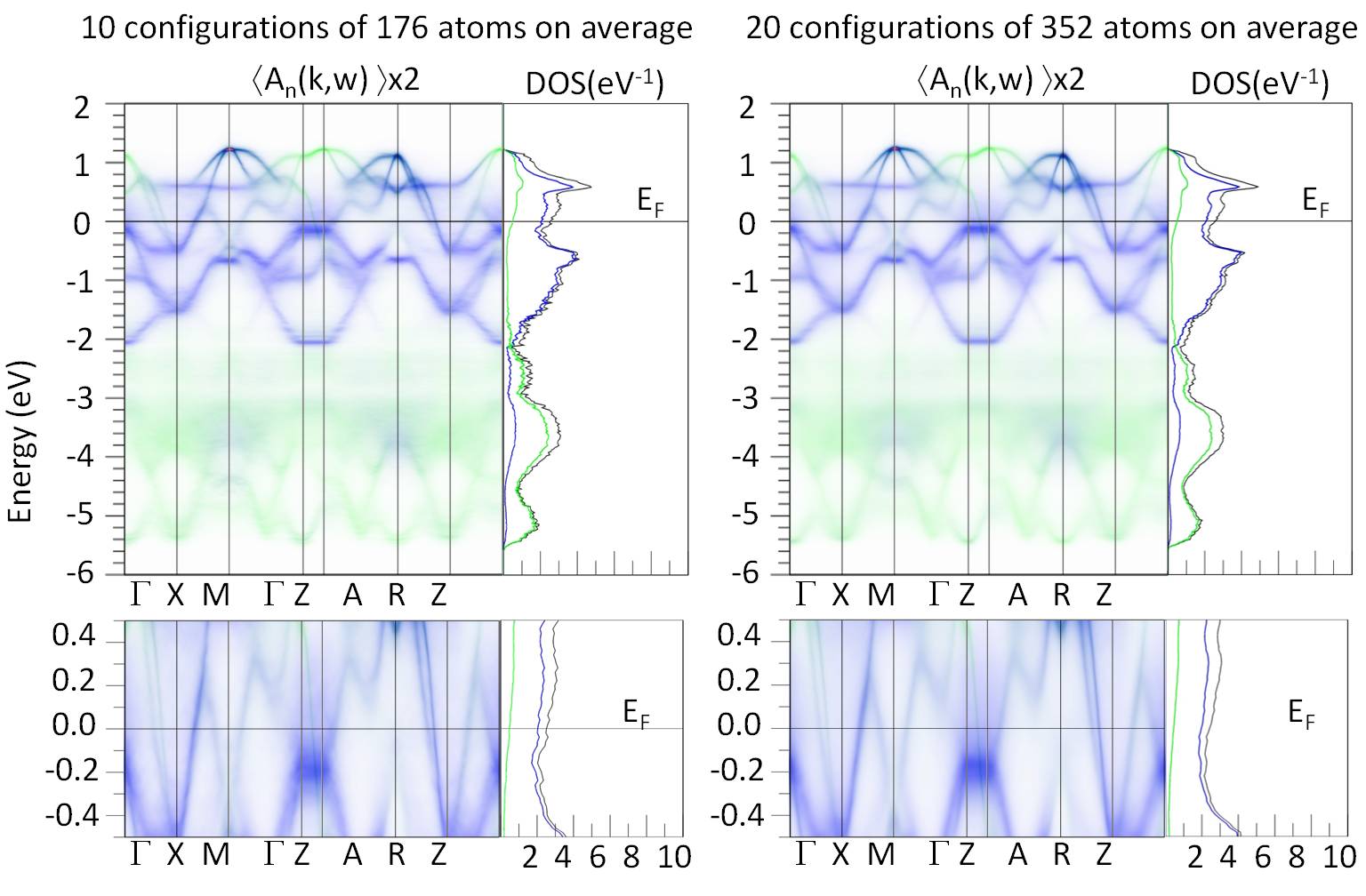}
\caption{\label{fig:figsup3}
}
\end{figure}

In figure \ref{fig:figsup3} we demonstrate the convergence of the spectral function of disordered K$_{0.8}$Fe$_{1.6}$Se$_{2}$ with respect to the number and the size of the configurations.

\end{widetext}

\end{document}